\documentclass[twocolumn]{aastex631}

\def\fedd{{\rm f}_{\rm Edd}}
\newcolumntype{C}{>{$}c<{$}}
\usepackage{amsmath}
\usepackage{enumitem}
\usepackage{float}

\newcommand{\MSun}{{\rm M}_{\sun}}
\newcommand{\dorb}{\delta \phi_{\rm orb}}
\newcommand{\dorbG}{\delta \phi^{\rm (GW)}_{\rm orb}}
\newcommand{\dorbN}{\delta \phi^{\rm (NoGW)}_{\rm orb}}
\newcommand{\dGW}{\delta\phi_{\rm GW}}

\begin{document}

\title{
{Gas-induced gravitational-wave dephasing and accretion periodicities of live post-Newtonian massive black hole binaries: warm disk}
}

\correspondingauthor{Mudit Garg}
\email{mg9113@nyu.edu}

\author[0000-0002-9032-9103]{Mudit Garg}
\affiliation{Center for Cosmology and Particle Physics, Physics Department, New York University, New York, NY 10003, USA}

\author[0000-0002-8400-0969]{Alessia Franchini}
\affiliation{Dipartimento di Fisica A. Pontremoli, Università degli Studi di Milano, Via Celoria 16, I-20134 Milano, Italy}

\author[0000-0001-6106-7821]{Alessandro Lupi}
\affiliation{DiSAT, Università degli Studi dell’Insubria, via Valleggio 11, I-22100 Como, Italy}
\affiliation{INFN, Sezione di Milano-Bicocca, Piazza della Scienza 3, I-20126 Milano, Italy}

\begin{abstract}

We perform 3D hydrodynamical simulations of an equal-mass quasi-circular live $10^6~\MSun$ massive black hole binary (MBHB) embedded in a prograde, locally isothermal circumbinary disk (CBD) with $0.1$ aspect ratio. The binary evolves under the effect of gaseous torques and $2.5$ post-Newtonian dynamics. {We follow the gravitational-wave (GW) driven inspiral from $53$ Schwarzschild radii ($r_s$) down to merger, over $1.5$ years.}
Comparing the GW inspiral rate with the viscous inflow, we find the binary to decouple from the CBD just $\approx2$ days before merger. We measure gas torques (gravitational and accretion) with and without concurrent GW emission, finding that their sums agree to within $\lesssim5\%$ down to $25~r_s$.
We then measure a gas-induced GW phase shift {that peaks at $\approx5\times10^{-4}$ rad near $45~r_s$, as the gas torque slows the inspiral, and then changes sign as the accreted mass accelerates it. It is LISA-detectable only at $z\lesssim0.1$}.
We further characterize the mass accretion rate modulations. {All three fundamental periodicities,} associated with the ``lump" at the inner cavity edge, {the binary orbital period, and half of it,} are shifted to lower frequencies {by precession}. We measure {the shift at the orbital period}, due to relativistic apsidal precession, in an inspiraling binary for the first time, and newly identify {the other two} as due to the precession of the eccentric cavity. Our results have implications for multi-messenger astronomy, since observation of accretion rate modulation by LSST/Roman surveys and phase-shift by LISA will provide crucial information on the complex environment surrounding MBHBs.

\end{abstract}


\section{Introduction} \label{S:Intro}

The adopted laser interferometer space antenna (LISA) mission \citep{Colpi2024}, together with the in-development TianQin \citep{Li2025} and Taiji \citep{Gong2021}, would start the era of milli-Hz gravitational wave (GW) astronomy in $2030$s. One of the primary sources of LISA is near-merger massive black hole binaries (MBHBs), made by two near equal-mass $\sim10^4$-$10^7~\MSun$ MBHs. LISA would be able to see MBHBs up to redshift $z\sim10$ due to their high signal-to-noise ratios (SNRs) of $\mathcal{O}(10^3)$. MBHBs will have hundreds to thousands of observable orbital cycles during their up to several years of inspiral before merger in the LISA band. This would allow us not only to constrain source properties but also to test our astrophysics \citep{AstroWG2023}, cosmology \citep{CosmoWG2023}, and fundamental physics \citep{FunWG2022} models in the strong gravity and at high-redshift.

The presence of MBHs at the center of most galaxies \citep{Kormendy2013} makes MBHBs a likely byproduct of galaxy mergers \citep{Begelman1980} in which, according to the hierarchical structure formation paradigm, smaller galaxies merge to form larger ones. During a galaxy merger, the two MBHs sink towards the center of the newly formed galaxy via dynamical friction (\citealt{Chandrasekhar1942}) against dark matter, gas, and stars, forming a bound MBHB on typical scales of a few pc. At this stage, dynamical friction becomes inefficient and the two-body relaxation time is longer than the Hubble time. This stalling of MBHBs has troubled theorists for several decades \citep{Milosavljevic2002}. However, better modeling of stellar distributions \citep{Preto2011,Khan2011,Vasiliev2015} and advances in understanding the binary-gas interaction \citep[see, e.g.,][]{MacFadyen2008, Mayer2013, Dittmann2022} have provided compelling evidence that MBHBs can sink further, reaching the GW-dominated regime on milli-pc scales over millions to billions of years but well within the lifetime of the Universe. After that, GWs merge the binary over a period of ten to a few hundred thousand years.

Because of their mass range, LISA MBHBs are more likely to reside in the gas-rich environments typical of Milky Way-like spiral galaxies. Gas can then lose angular momentum due to gravitational instabilities or supernova feedback \citep{AstroWG2023} and fall to the galactic nucleus to interact with MBHBs. Within the MBHB potential well, gas tends to settle in a likely co-planar circumbinary disk (CBD; \citealt{Escala2004, MacFadyen2008}) whose thickness is determined by the efficiency of radiative cooling \citep{Shakura1973}. The gravitational torque provided by the CBD on the binary can shrink the MBHB within a $100$ Myr \citep{Haiman2009} for most binary-disk parameters, as shown by numerous hydrodynamical (HD) studies in the last decade \citep{DOrazio2016,Zrake2021,DOrazio2021,Tiede2020,Tiede2024,Franchini2021,Franchini2022,Dittmann2022,Siwek2023}.

Another recent breakthrough has been the identification of the exact moment when the MBHB decouples from CBD, once GWs become the dominant binary evolutionary mechanism. Contrary to the original idea of the decoupling occurring when the GW-driven merger timescale \citep{Peters1964} becomes shorter than the viscous time \citep[see, e.g.,][]{ArmitageNatarajan2002}, recent works \citep{Dittmann2023, ONeill2025} have argued that the actual decoupling happens only a few days before merger. In particular, this result was achieved by means of 2D simulations of an evolving gas flow around a fixed Newtonian MBHB which was shrunk on a pre-determined GW timescale \citep{Peters1964}, after an initial relaxation phase in which some sort of binary-disk steady state at a given separation was reached. 

The fact that CBDs can follow the shrinking MBHBs in the LISA band opens up the possibility that LISA, with its high sensitivity, could also observe gas imprints on the gravitational waveform. The characterization of environmental effects on LISA sources has been active in recent years, especially for the highly unequal-mass LISA sources \citep{Derdzinski2021,Zwick2022,Garg2022,Speri2023,Duque2025,Garg2026,Garg2026c}, as their $\lesssim\mathcal{O}(10^5)$ GW cycles make it easy to detect small perturbations. Although MBHBs spend relatively fewer cycles in the LISA band, their SNRs are much higher, hence they could represent another important probe of gas physics \citep{Barausse2014,Garg2022,Garg2024b,Garg2024c,Garg2024d,Zwick2024,Zwick2025}. A fundamental limitation of almost all the above-mentioned works is that they measure gas imprints by linearly adding the semi-analytically modeled semi-major axis change rate due to just gas, informed by non-GW driven, purely HD simulations, to the GW-driven semi-major axis rate of change in order to compute the orbital phase-shift. While efficient in terms of computational cost, these models might miss the coupling between GW and gas. Hence, it is necessary to run 3D HD simulations of live binaries that evolve via GW emission to investigate the phase shift precisely.

In \citet{Garg2025} (hereafter Paper I), we investigated the effect of dynamically coupling GW and gas torques on the evolution of MBHB a year away from its merger, i.e.~in the LISA band, by means of 3D, live-binary, post-Newtonian (PN) HD simulations of a $0.03$ aspect ratio CBD. We reported that the mean gas torque is $\sim20\%$ weaker in the case of GW+gas as compared to the gas-only simulation, when compared over the same integrated $278$ orbits, most likely due to rapid binary shrinkage via GWs. We used $2.5$PN simulations to measure the coupled effect on the binary orbit, finding a gas-induced phase-shift $\sim7$ times smaller than the semi-analytical expectation.

In this work, we extend the parameter space by performing numerical simulations of a CBD with aspect ratio $H/R=0.1$ ($H$ being the disk scale height and $R$ the radius) surrounding an equal-mass live binary that evolves via GW radiation and gaseous torques. {The aspect ratio of CBDs around LISA MBHBs is uncertain. A recent radiation-magnetohydrodynamic study of a binary at a separation comparable to ours, accreting at a rate a few times smaller than here, finds $H/R\sim0.03$, about half as thick as without radiation \citep{Tiwari2025}. Since in the radiation-pressure dominated regime the scale height grows linearly with the accretion rate \citep{Shakura1973}, our previous ($H/R=0.03$; cold disk) and current ($H/R=0.1$; warm disk) studies bracket the plausible range, and also lie on either side of the transition between disks that extract and disks that add angular momentum to the binary \citep{Tiede2020,Dittmann2022,Franchini2022}. Furthermore, $H/R=0.1$} is by far the most studied case, as larger aspect ratios lead to a shorter viscous time, therefore reducing the computational cost \citep[see, e.g.,][]{Duffell2024}{, which allows us to validate our torques against the literature.}

\section{Numerical setup} \label{S:setup}

Similar to the approach taken in Paper I, we adopt the setup from \citet{Franchini2024} for $H/R=0.1$. We model the $M=10^6~\MSun$ total mass MBHB with two Schwarzschild MBHs represented by two equal-mass sink particles with their radius being the innermost stable circular orbit of the respective MBH. The binary is quasi-circular{, with a small residual initial eccentricity $e\approx0.024$}, with initial semi-major axis ($a$) ${53}~r_s$, where $r_s\equiv2GM/c^2$ is the Schwarzschild radius of the binary. The initial semi-major axis is {about} twice the classical\footnote{Based upon timescales comparison between the binary inspiral rate and the disk's local viscous inflow timescale at the inner cavity edge.} gas-binary decoupling radius {$a^{\rm cl}_{\rm dec}\sim 25r_s$} \citep{Milosavljevic2005,Tang2018}, and corresponds to the typical separation at which this binary will enter the LISA band at redshift $z\sim1$. 

The initial conditions are taken from \citet{Franchini2022} live binary GIZMO \citep{Hopkins2015} simulations, where a 3D, locally isothermal, Newtonian, thin disk with $H/R=0.1$, and Shakura–Sunyaev \citep{Shakura1973} viscosity coefficient $\alpha=0.1$ is evolved for $1000$ orbital periods ($P_{\rm B}$) to reach a steady state. The disk had initially $2\times10^6$ gas particles sampled from a surface density profile $\Sigma\propto R^{-3/2}$ from $R=2a$ to $10a$. After $1000$ orbits, the disk outer edge spreads to $20a$, the disk inner edge increased to $3a$, the cavity becomes eccentric and gas forms a minidisk around each sink. Unlike \citet{Franchini2024}, where the CBD mass was $M_{\rm d}=100\MSun$, in this work we employ a CBD of $0.06~\MSun$ to keep the Eddington rate\footnote{$\fedd=\dot M/\dot M_{\rm Edd}$, where $\dot M$ is the ratio of the gas accretion rate onto the binary and $\dot M_{\rm Edd}=M/50$ Myr for our $10\%$ assumed radiative efficiency.} $\fedd$ close to unity. We take the sound speed $c_s$ profile from \citet{Farris2014}, such that the kinematic viscosity $\nu\equiv\alpha c_s H={1.8\times10^{-3}}$ in code units (i.e., $G=M=a=1$) at $R=3a$.

Analogously to Paper I, we employ four GIZMO simulation setups: gas+$2.5$PN, gas+$2$PN, $2.5$PN, and $2$PN. Here, the keyword ``gas" represents the presence of a CBD, ``$2.5$PN" means that each MBH is dynamically evolved \citep{Blanchet2014} with conservative $1$PN and $2$PN terms, and radiative $2.5$PN terms on top of the Newtonian effect. The $2.5$PN term results in the GW-driven inspiral (see \citealt{Franchini2024} for technical details). Lastly, the ``$2$PN" keyword refers to simulations in which we neglect GW emission, i.e.~the $2.5$PN term. The gas+$2$PN and $2$PN are performed to isolate the effect of GW on both gas torques onto the binary and the gas-induced orbital phase-shift, computed later during post-processing in \S~\ref{Sec:Results}. The $2.5$PN and $2$PN runs are performed to compute phase-shifts relative to their hydro counterparts. Similarly to Paper I, we perform a resolution study for the gas+$2.5$PN case by increasing the number of splitting levels in the hyper-Lagrangian refinement, evolving every case for $100$ orbits. Expressed in terms of the inter-particle spacing $\Delta x$ at $3a$, we have the low-resolution run (LR) with $\Delta x[3a]={3.6~r_s}$, the mid-resolution (MR) with $\Delta x[3a]={2.1~r_s}$, and the high-resolution (HR) with $\Delta x[3a]={1.2~r_s}$. We find that the MR run is sufficiently converged in terms of gas torque and phase shift (see Appendix~\ref{App:Res}). We therefore report the results obtained using the MR resolution simulation, unless otherwise stated.

We evolve the gas+$2.5$PN setup {until the MBHB merges, i.e.~until the black holes are separated by $\sim4~r_s$. The binary orbital period $P_{\rm B}$ decreases significantly, from $\approx9.75$ hours to $\approx0.30$ hours. We also evolve the gas+$2$PN, the $2.5$PN and the $2$PN setups over the same integration time. The two MBH sink particles are integrated with the same timestep ($\approx2.2$~s) in all simulations, to allow for a proper comparison of the binary evolution with and without the presence of the disk. {Since the gas runs use a coarser MBH timestep during their first orbit, each vacuum run starts from the binary of its gas run after this orbit.} 

In total, the simulations (including the resolution study) used $\sim10^{5}$ CPU core hours with main contributions from the gas+$2.5$PN run ($\sim40,000$ hours) and the gas+$2$PN run ($\sim50,000$ hours), where the former becomes faster during the binary inspiral.}

\section{Post-processing analysis}\label{Sec:Postprocess}
The CBD exerts both a gravitational torque ($T_{\rm grav}$) and an accretion torque ($T_{\rm acc}$) onto the binary. The former originates from non-axisymmetric features in the gas flow, while the latter is caused by accretion of particles onto the MBHs. When a gas particle enters the sink radius, it is flagged for accretion and gets promptly (within the next coarse step) removed from the domain, while its mass and angular momentum are consistently added to the sink \citep{Bate1995,Price2018,Franchini2021}.

We compute three dimensionless quantities:
\begin{align}
    \xi_{\rm acc}&=\frac{T_{\rm acc}}{\dot M a^2\Omega_{\rm B}},\\
    \xi_{\rm grav}&=\frac{T_{\rm grav}}{\dot M a^2\Omega_{\rm B}},\nonumber\\
    \xi_{\rm grav+acc}&=\frac{T_{\rm grav}+T_{\rm acc}}{\dot M a^2\Omega_{\rm B}}\nonumber,
\end{align}
where $\Omega_{\rm B}\equiv\sqrt{GM/a^3}=2\pi/ P_{\rm B}$ is the binary orbital angular frequency. The eigenvalue $\xi_{\rm grav+acc}$ is similar to the accretion eigenvalue used in \cite{Duffell2024}. The accretion and gravitational eigenvalues depend sensitively on the binary parameters (especially binary mass ratio and eccentricity $e$), dimensionality of the setup (i.e.~2D/3D), whether the binary is prograde/retrograde, and on the gaseous disk thermodynamics (e.g. the disk aspect ratio $H/R$). We further break down $\xi_{\rm grav}$ into the gravitational torque from particles outside the binary orbit $R>a$ ($\xi_{R>a}$) and inside the binary orbit $R<a$ ($\xi_{R<a}$) to isolate the effect of the outer CBD from the effect of the two minidisks. Similarly to Paper I, thanks to our live binary evolution, we can directly measure different $\xi$ and $\fedd$ directly from our CBD simulations gas+$2.5$PN and gas+$2$PN.

{We also estimate the MBHB semi-major axis $a_{\rm dec}$ at which the CBD decouples from the binary in the gas+$2.5$PN run using a velocity-based criterion (hence, different than $a^{\rm cl}_{\rm dec}$) that compares the binary's inspiral rate directly to the local viscous inflow velocity at the cavity wall \citep{ArmitageNatarajan2002}.  The binary and disk are considered coupled while
\begin{equation}\label{eq:adec}
    v_r(a_{\rm cav})= \frac{3\nu(a_{\rm cav})}{2a_{\rm cav}} > |\dot a|~\forall ~a>a_{\rm dec},
\end{equation}
and decoupled once the binary's inspiral outpaces the rate at which the disk can viscously resupply the cavity edge. The quantity $v_r(a_{\rm cav})$ is the viscous velocity evaluated at the cavity semi-major axis $a_{\rm cav}$, computed from the peak of the gas mass distribution, which is binned by each fluid element own osculating semi-major axis \citep{Dittmann2023}.}

The gaseous CBD can expand or shrink the binary, leading to more or fewer binary cycles compared with its evolution in vacuum. This difference manifests as a phase shift $\dorb$. Analogously to Paper I, since our live-binary setup allows us to measure the {orbital} phase directly, we can compare {the simulations with and without the gaseous disc}
to compute the gas-induced phase shifts with ($\dorbG$) and without ($\dorbN$) GW emission. {$\dorbN$ isolates the effect of the gas alone, while $\dorbG$ captures the dynamically coupled evolution under both GW emission and the CBD.}

For our quasi-circular binary, the GW phase-shift $\dGW$ is simply $2\dorbG$. {We measure $\delta \phi^{\rm (GW)}_{\rm orb,\langle a\rangle}$ by computing the difference between the orbital phase of the gas+$2.5$PN and $2.5$PN runs at the same (orbit-averaged) separation $\langle a\rangle$, accumulated from the initial to the decoupling radius.

We orbit-averaged the semi-major axis $a$ as it oscillates within each orbit under the non-Keplerian (PN and gas) forces. We compare at equal separation rather than equal time because the two runs inspiral at slightly different rates.

We also measure the apsidal phase-shift $\delta\phi^{\rm(GW)}_{\dot\varpi_{\rm gas,\langle a\rangle}}$ due to the disk-induced binary precession rate $\dot\varpi_{\rm gas}$ \citep[see, e.g.,][]{Tiede2024b}.

Since the semi-major axis of the \{gas+$2$PN, $2$PN\} pair does not significantly evolve, we also compare the effect of the gaseous disk with and without concurrent GW emission at fixed time, computing $\delta\phi^{\rm (GW)}_{\rm orb,t}$ and $\delta\phi^{\rm (NoGW)}_{\rm orb,t}$ over the same $500$ initial binary orbits of the gas+$2$PN run for fair comparison.}

\section{Results}\label{Sec:Results}

{We first estimate the decoupling radius $a_{\rm dec}$ using the velocity-based criterion expressed in Eq.~\ref{eq:adec}. The measured initial value of the cavity inner edge is $a_{\rm cav}\simeq 3a$, broadly consistent with other simulations of circular binaries embedded in accretion disks. 
We find $a_{\rm dec}\approx13~r_s$, which is about half of the classical decoupling radius $25~r_s$ \citep{Milosavljevic2005} and consistent with the estimate reported in \citet{Dittmann2023}. During inspiral, the ratio $a_{\rm cav}/a$ begins to depart from its plateau value of $\sim3$ at $a\sim27~r_s$, rising to $\sim9$ at $a_{\rm dec}$. {The crossing of the binary inspiral speed and the viscous inflow speed at the cavity edge are shown in Fig.~\ref{fig:fEdd}.}}

\begin{figure*}
    \centering
    \includegraphics[width=0.9\linewidth]{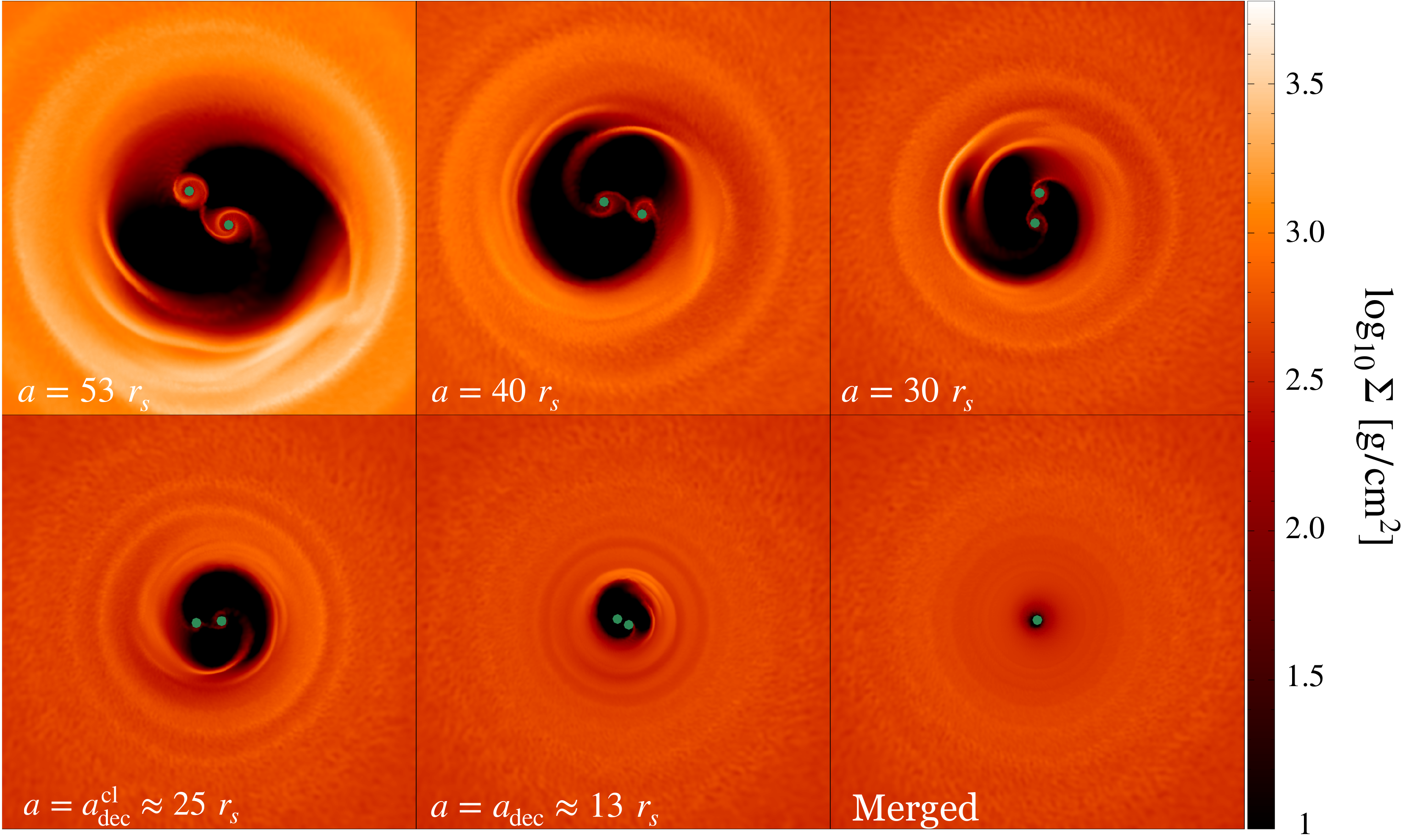}
    \caption{Column density ($\Sigma$) plots at six semi-major axes: $a={53}~r_s$ (top-left), $a={40}~r_s$ (top-center), {$a=30~r_s$ (top-right), $a=a^{\rm cl}_{\rm dec}\approx25~r_s$ (classical decoupling radius; bottom-left), $a=a_{\rm dec}\approx13~r_s$ (decoupling radius; bottom-center), and post-merger (bottom-right)} for the binary evolution under both GW and gas varying from $\sim10$ to $6\times10^3$ g/cm$^{2}$. The plots cover a region $[-5a,5a]$ {based on the first panel}. Similar to \citet{Franchini2024}, both the binary and the cavity shrink with time.}
    \label{fig:disk}
\end{figure*}

We show the disk surface density in Fig.~\ref{fig:disk} for our gas+$2.5$PN MR simulation at {different binary} semi-major axes: $a={53}~r_s$, $a={40}~r_s$, {$a=30~r_s$, $a=a^{\rm cl}_{\rm dec}\approx25~r_s$, $a=a_{\rm dec}\approx13~r_s$, and post-merger}. {At $a=a^{\rm cl}_{\rm dec}$, a small amount of gas is still being supplied to each mini-disk, while at $a=a_{\rm dec}$ there is no more supply and one mini-disk has already been depleted}. The morphologies {in the top three panels (i.e.~$a\geq30r_s$) expectedly look similar to the} GW-driven inspiral snapshots {reported in} \citet{Franchini2024}, where both the binary and the cavity shrink with time via GW emission. In contrast to Paper I with $H/R=0.03$, the minidisks here {remain denser until approximately the decoupling radius}.
This is consistent with the picture that thicker disk allows a larger amount of material to leak into the cavity \citep{Ragusa2016,Franchini2022}.

\begin{figure}
    \centering
    \includegraphics[width=0.5\textwidth]{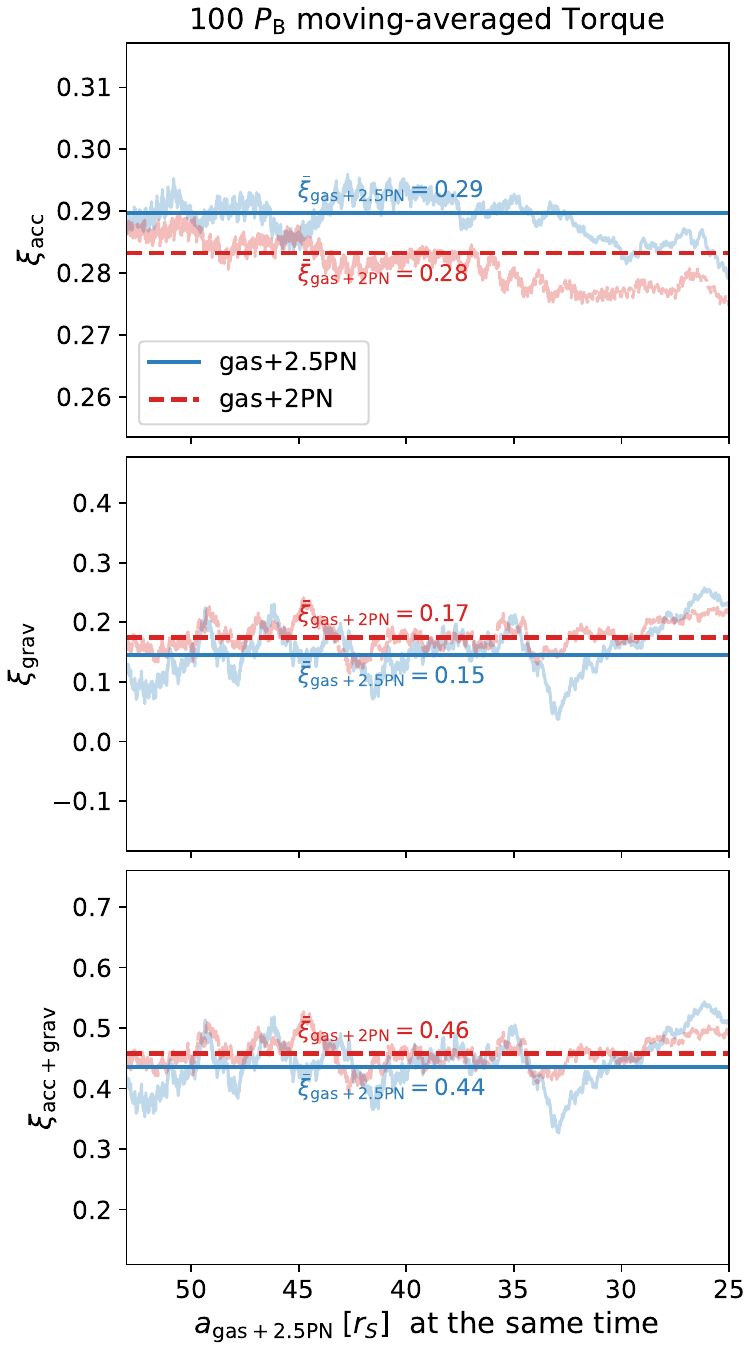}
    \caption{$100-$orbit moving time-average of accretion torque ($\xi_{\rm acc}$; top panel), gravitational torque ($\xi_{\rm grav}$; middle panel), and their sum ($\xi_{\rm acc+grav}$; lower panel) between the start of the simulation until {$a_{{\rm gas}+2.5{\rm PN}}=25~r_s$} for both MR gas+$2.5$PN (solid blue line) and MR gas+$2$PN (red dashed line) setups. For each panel, we also show respective mean value for each curve over the whole duration. Here $\{\bar\xi_{\rm acc},\bar\xi_{\rm grav},\bar\xi_{\rm acc+grav}\}$ are $\sim\{{0.29,0.15,0.44}\}$ for the GW+gas run and $\sim\{{0.28,0.17,0.46}\}$ for the gas-only setup. During the long evolution, $\bar\xi_{\rm acc}$ in the GW run is slightly higher than that in its non-GW counterpart. The opposite occurs for $\bar\xi_{\rm grav}$, where its values are higher for the gas+$2$PN simulation. Overall, $\bar\xi_{\rm acc+grav}$ is slightly higher for the non-GW setup.}
    \label{fig:Gastorque}
\end{figure}

Fig.~\ref{fig:Gastorque} shows the $100$-orbit moving averaged values of the dimensionless gas torque parameters $\xi_{\rm acc}$, $\xi_{\rm grav}$, and $\xi_{\rm acc+grav}$ for our CBD runs gas+$2.5$PN and gas+$2$PN {as a function of
the semi-major axis of the former, $a_{{\rm gas}+2.5{\rm PN}}$, with both runs compared at the same time. We show results only down to $a_{{\rm gas}+2.5{\rm PN}}=25~r_s$, since below it the CBD progressively decouples from the binary and its torques become dynamically unimportant compared with GW emission. The horizontal lines show the mean of each quantity over the whole range.} We find $\bar\xi_{\rm acc}\sim {0.29},~\bar\xi_{\rm grav}\sim {0.15},~\bar\xi_{\rm acc+grav}\sim {0.44}$ for the GW+gas case and $\bar\xi_{\rm acc} \sim {0.28},~\bar\xi_{\rm grav}\sim {0.17},~\bar\xi_{\rm acc+grav}\sim{0.46}$ for the gas-only simulation. We can see that the values of the dimensionless parameter $\xi$ oscillate around the mean value over time. Any attempt to deduce physical meaning from these oscillations would need a careful assessment of the level of particle noise within the simulation. We therefore refrain from drawing conclusions about the shape of these curves.

\begin{figure}
    \centering
    \includegraphics[width=0.5\textwidth]{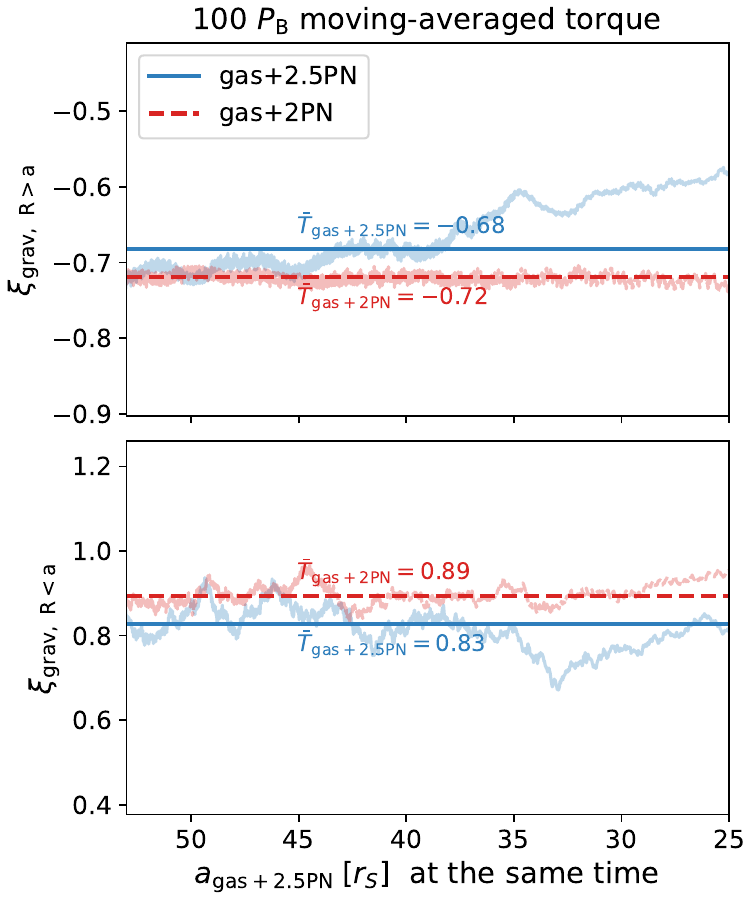}
    \caption{Same as in Fig.~\ref{fig:Gastorque} but for the excised gravitational torque ($\xi_{\rm grav,R>a}$; top panel) and the minidisk gravitational torque ($\xi_{\rm grav,R<a}$; bottom panel). Here the overall averages $\{\bar\xi_{\rm grav,R<a},\bar\xi_{\rm grav,R>a}\}$ are $\{{0.83,-0.68}\}$ for the gas+$2.5$PN simulation and $\sim\{{0.89,-0.72}\}$ for the gas+$2$PN setup.}
    \label{fig:xigrav}
\end{figure}

In order to better investigate the behavior of the gravitational torque, in Fig.~\ref{fig:xigrav} we show the breakdown of the torque coming from the excised region ($\bar\xi_{\rm grav,R>a}$; from gas particles $R>a$) and from the minidisks region ($\bar\xi_{\rm grav,R<a}$; from gas particles $R<a$) by again applying the $100-$orbit moving average. We find $\bar\xi_{\rm grav,{R>a}}\sim {-0.68},~\bar\xi_{\rm grav,{R<a}}\sim {0.83}$ for the GW+gas simulation and $\bar\xi_{\rm grav,{R>a}}\sim {-0.72},~\bar\xi_{\rm grav,{R<a}}\sim {0.89}$ for the gas-only run. 
We find, consistently with previous works in the literature, that the excised region extracts angular momentum from the binary while the region inside the binary orbit exerts a positive torque, therefore widening the binary orbit for the non-GW case and slowing down the inspiral for the GW+gas simulation. 
{Since $\xi$ is normalized by the accretion rate, a reduced gas supply lowers both the torque and $\dot M$ and leaves $\xi$ nearly unchanged. The progressive decoupling of the disk is instead visible in the accretion rate (Fig.~\ref{fig:fEdd}).}

We measure the Eddington ratio $\fedd$ until $25~r_s$ for both runs (see Fig.~\ref{fig:fEdd}) and find the time-averaged value $\bar{\rm f}_{\rm Edd}\sim 0.55$ to be similar for both the GW and non-GW runs, the former having a slightly lower value because of the progressive decoupling of the CBD.

\subsection{Accretion periodicities}\label{S:Mdot}

\begin{figure*}
    \centering
    \includegraphics[width=\textwidth]{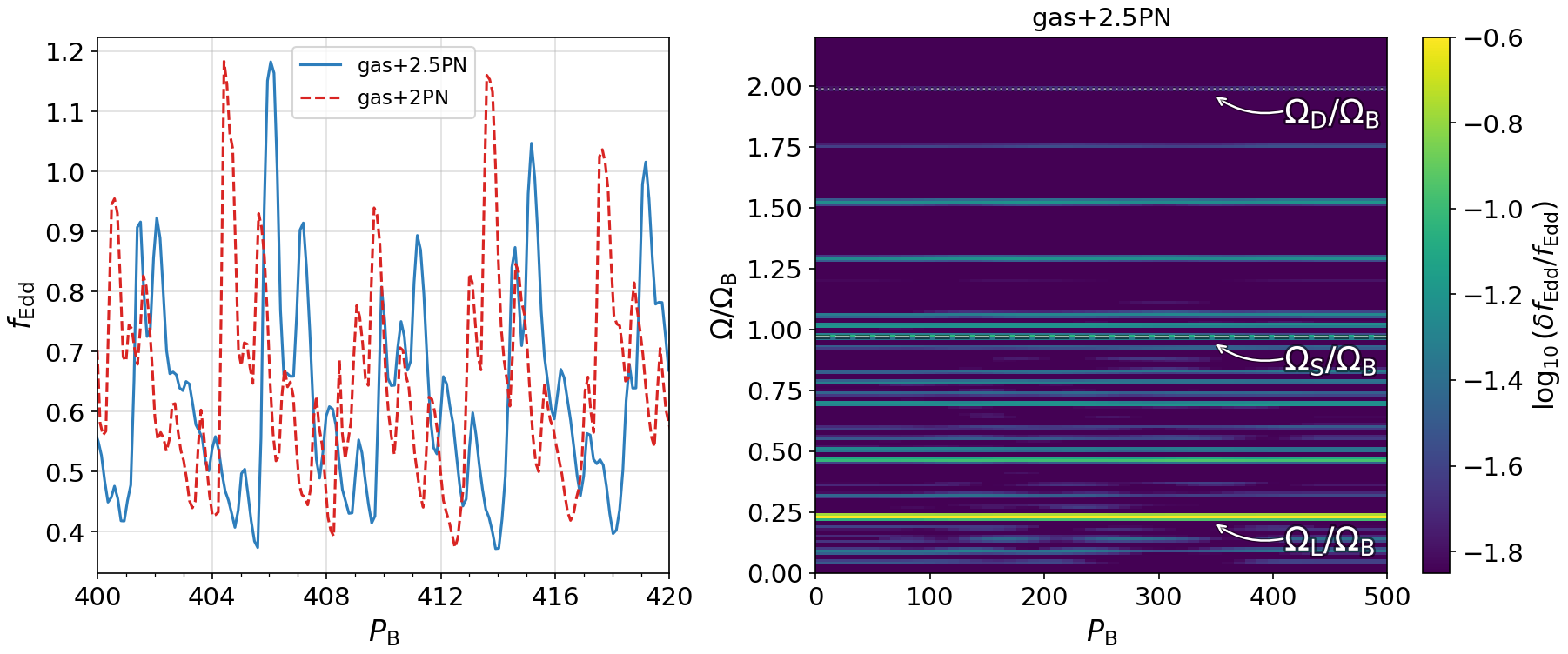}
    \caption{Left: Eddington fraction $\fedd$ evolution as a function of time in binary cycles between ${400}~P_{\rm B}$ and ${420}~P_{\rm B}$ {for both gas+$2.5$PN (solid blue) and gas+$2$PN (dashed red) runs.} Right: spectrogram of {$\fedd$, normalized by its running mean over $100$ orbits (frequency resolution $0.01~\Omega_{\rm B}$), at different frequencies (normalized by ${\Omega}_{\rm B}$) in the first $500~P_{\rm B}$ for the gas+$2.5$PN run. The colorbar shows $\log_{10}(\delta\fedd/\fedd)$, the fractional amplitude of the $\fedd$ modulation at each frequency. The spectrum is set by three frequencies: $\Omega_{\rm L}\approx0.233~\Omega_{\rm B}$ (initial $\delta\fedd\approx0.25$), $\Omega_{\rm S}\approx0.97~\Omega_{\rm B}$ (initial $\delta\fedd\approx0.090$), and $\Omega_{\rm D}\approx1.986~\Omega_{\rm B}$ (initial $\delta\fedd\approx0.018$); the latter two are marked by the white dashed and dotted lines. The remaining strong lines are harmonics and linear combinations of these three frequencies (see Fig.~\ref{fig:Periodogram}). The color-scale floor at $10^{-1.85}$ lies just below the weakest fundamental line $\Omega_{\rm D}$.}}
    \label{fig:fEddMR}
\end{figure*}

We show in Fig.~\ref{fig:fEddMR} the accretion rate time series {(left panel)}, in terms of the Eddington ratio $\fedd$, between ${400}~P_{\rm B}$ and ${420}~P_{\rm B}$ for both CBD simulations. We show these time series for only 20 orbits in order to easily distinguish the two light curves with or without GWs.  

We compute the spectrograms of $\fedd$ over $500$ binary cycles. We use the orbital phase as the time coordinate to remove the imprint of the evolving binary orbital period. In units of $\Omega_{\rm B}$, all lines therefore remain nearly stationary, except for the small drift of $\Omega_{\rm S}$ discussed below. We only show the spectrogram of the gas+$2.5$PN run in Fig.~\ref{fig:fEddMR}, as that of the gas+$2$PN run is very similar (see also Fig.~\ref{fig:Spec_initial}).

{Four frequencies are relevant here. The binary orbits at $\Omega_{\rm B}$, and its mildly eccentric orbit undergoes relativistic periapsis precession at a rate $\dot\varpi$, so that the binary separation oscillates at the radial frequency $\Omega_{\rm B}-\dot\varpi$. The eccentric cavity precesses at a much slower rate $\Omega_p$, and the overdense lump at the cavity edge orbits at $\Omega_{\rm lump}$. The total accretion rate does not depend on any fixed direction in space, so it can only be modulated at the frequency of one of these structures relative to another.}

All the horizontal lines in the spectrum are integer combinations $n_1\Omega_{\rm L}+n_2\Omega_{\rm S}+n_3\Omega_{\rm D}$ of three fundamental frequencies (see Fig.~\ref{fig:Periodogram} for details). 
The first, $\Omega_{\rm L}\approx0.233~\Omega_{\rm B}$ (a period of $\approx4.3~P_{\rm B}$), is {produced by} the overdense lump \citep[see, e.g.,][]{MacFadyen2008,Shi2012,Noble2012} orbiting at the cavity inner edge, as widely reported in the literature \citep[see, e.g.,][]{Lai2023,Duffell2024,MignonRisse2025}{, although, as we show below, it is not the lump orbital frequency itself}. Note that the lump is visible in the first panel of Fig.~\ref{fig:disk}. Neither its frequency nor its fractional amplitude $\delta\fedd/\fedd$ shows any clear trend with the GW-driven inspiral over the $500$ binary cycles of Fig.~\ref{fig:fEddMR}.

The second, $\Omega_{\rm S}=\Omega_{\rm B} -\dot{\varpi}\approx0.97~\Omega_{\rm B}$, is the radial frequency of the mildly eccentric binary, shifted below $\Omega_{\rm B}$ by\footnote{$\dot{\varpi}$ is measured from the phase of the orbit-averaged eccentricity vector and matches the $1$PN periapsis advance, i.e.~$6\pi GM/[c^2a(1-e^2)]$ per orbit, within $1\%$.} $\dot{\varpi}\approx3.0\times10^{-2}~\Omega_{\rm B}$. The frequency $\Omega_{\rm S}$ drifts with the GW-driven inspiral, from $0.971$ to $0.969~\Omega_{\rm B}$ over the $500$ orbits of Fig.~\ref{fig:fEddMR}, since $\dot\varpi/\Omega_{\rm B}$ increases as the binary shrinks, while the fractional amplitude of this line, which scales linearly with the eccentricity, decreases from $\delta\fedd/\fedd\approx0.090$ to $0.077$ as GWs circularize the binary from $e\approx0.024$ to $0.021$. This behavior persists along the inspiral at least down to $\sim37~r_s$.

The third, $\Omega_{\rm D}\approx1.986~\Omega_{\rm B}$, is consistent with the twice-per-orbit modulation produced as the two MBHs alternately pull streams from the cavity edge. { We measure} $\Omega_p\approx7.01\times10^{-3}~\Omega_{\rm B}$.\footnote{We adopt the per-element cavity eccentricity vector definition of \citet{Duffell2024} and compute $\Omega_p$ from the change in its unwrapped phase per unit orbit. Our measured $\Omega_p$ is consistent with \citet{Duffell2024} results for 3D simulations.} The MBHs therefore pass near the cavity edge at the relative frequency $\Omega_{\rm B}-\Omega_p$, giving a modulation on the frequency $2(\Omega_{\rm B}-\Omega_p)\approx1.986~\Omega_{\rm B}$. There is no measurable impact of $\dot\varpi$ on $\Omega_{\rm D}$: the former sets the rate at which the separation between the two MBHs oscillates, while the latter depends on the angle between the line joining the two MBHs and the cavity eccentricity vector. Accordingly, $\Omega_{\rm D}$ matches $2(\Omega_{\rm B}-\Omega_p)$ to $10^{-5}~\Omega_{\rm B}$, over the window of Fig.~\ref{fig:fEddMR}, while the harmonic at $2(\Omega_{\rm B}-\dot\varpi)$ is about $5$ times weaker and falls below our floor. Unlike $\Omega_{\rm S}$, $\Omega_{\rm D}$ does not drift with the inspiral, and it traces the cavity precession rate to within $3\times10^{-4}~\Omega_{\rm B}$ down to $32~r_s$, while its fractional amplitude remains nearly constant.

{The lump modulation is also shifted by the cavity precession. In the periodogram (Fig.~\ref{fig:Periodogram}), all lump combination lines appear at $\Omega_{\rm D}-n\Omega_{\rm L}=2(\Omega_{\rm B}-\Omega_p)-n\Omega_{\rm L}$ ($n=1$--$7$), and none at $2\Omega_{\rm B}-n\Omega_{\rm L}$, so the binary interacts with the lump at its frequency relative to the precessing cavity. The pattern speed of the $m=1$ density mode at the cavity edge, which we measure from our snapshots,\footnote{ From the Fourier transform of the $m=1$ mode of the gas at $2$--$3.5a$, using snapshots every $0.5~P_{\rm B}$ over $\sim1400$ orbits of the gas+$2$PN run ($\sim10^{-3}~\Omega_{\rm B}$ resolution).} is $\Omega_{\rm lump}\approx0.240~\Omega_{\rm B}=\Omega_{\rm L}+\Omega_p$, corresponding to a Keplerian orbit at $\approx2.6a$. The lump line $\Omega_{\rm L}=\Omega_{\rm lump}-\Omega_p$ is thus the lump frequency relative to the cavity, and all three fundamental frequencies are shifted by precession: $\Omega_{\rm S}$ by that of the binary, $\Omega_{\rm L}$ and $\Omega_{\rm D}$ by that of the cavity.}

We note that \citet{DeLaurentiis2025} also reported a related shift of the modulation frequency measured from simulations where the binary follows a fixed 2PN quasi-Keplerian orbit. For their $e=0.45$, $a=175~r_s$ binary, the accretion rate peaks at the radial period rather than the orbital period, which they attribute to apsidal precession, though they note the two periods are nearly indistinguishable in their spectra. Their precession of $3.97^{\circ}$ per orbit puts the radial period at $1.011~P_{\rm B}$, i.e.~$\Omega_{\rm S}=0.989~\Omega_{\rm B}$, a fixed value, whereas in our live binary $\Omega_{\rm S}$ evolves with the inspiral. In contrast, the {shifts of the lump and twice-per-orbit modulations} by the precession of the eccentric cavity, i.e.~{$\Omega_{\rm L}=\Omega_{\rm lump}-\Omega_p$ and} $\Omega_{\rm D}=2(\Omega_{\rm B}- \Omega_p)$, {have} not been reported or measured before.

\subsection{Phase-shift}\label{Sec:dphi}

\begin{figure}
    \centering
    \includegraphics[width=0.5\textwidth]{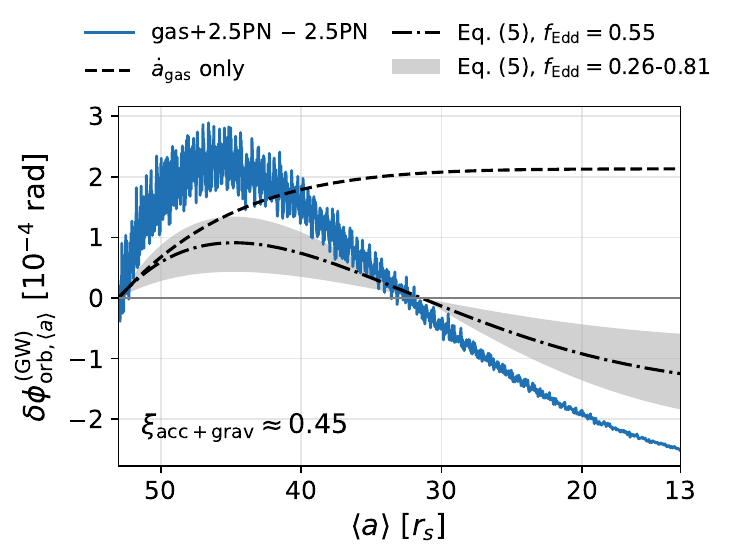}
    \caption{Gas-induced orbital phase shift at fixed $\langle a\rangle$ (blue), compared with Eq.~\ref{Eq:analytical} for $\xi_{\rm acc+grav}=0.45$ and $\fedd=0.55$ (dash-dotted) or $0.26$--$0.81$, the range in our simulation (gray band), and with only its $\dot a_{\rm gas}$ term (dashed). The small oscillations are due to the apsidal precession of the binary.}
    \label{fig:dphi_time}
\end{figure}

Similarly to Paper I, we numerically compute {the phase shifts defined in \S~\ref{Sec:Postprocess}. We first compare the GW-driven binary pair $\{{\rm gas}+2.5{\rm PN},~2.5{\rm PN}\}$ from $\langle a\rangle=53~r_s$ to our measured decoupling radius $\langle a_{\rm dec}\rangle=13~r_s$. Over this range, the presence of the gaseous disk causes a shift in the orbital phase {that first grows to $\delta\phi^{\rm (GW)}_{\rm orb,\langle a\rangle}\approx+2.5\times10^{-4}$~rad near $\langle a\rangle\approx45~r_s$, after $\approx700$ orbits, and then decreases, changing sign near $32~r_s$ and reaching $\approx-2.4\times10^{-4}$~rad at $a_{\rm dec}$ (Fig.~\ref{fig:dphi_time}). The corresponding GW phase shift is therefore $|\delta\phi_{\rm GW,\langle a\rangle}|\lesssim5\times10^{-4}$~rad. The peak is $\approx5.7\times10^{-8}$ of the orbital phase accumulated until $45~r_s$.\footnote{This is comparable to the disk-to-binary mass ratio $M_{\rm d}/M\approx6\times10^{-8}$, which sets the relative strength of the gas force perturbing the binary.} The phase shift turns over because accretion has two opposite effects. The positive gas torque slows the inspiral and dominates at large separations. The accreted mass instead makes the binary emit GWs and orbit faster at a given separation{, an effect set by the total mass accreted so far rather than by the current accretion rate. It therefore keeps growing as the accretion rate and the torque drop (Fig.~\ref{fig:fEdd}) and eventually dominates} (see below). The time of coalescence changes by $|\Delta t_{\rm c,gas}|\lesssim1$~s, within our timing resolution. In the GW signal, the phase shift is measured at fixed GW frequency rather than at fixed $\langle a\rangle$. The two are equivalent unless the gas also changes the orbital frequency at a given separation. We find this change to be so small, a fraction $\sim10^{-8}$, that the two comparisons differ by {at most $\approx2.5\times10^{-4}$~rad,} so that $|\delta\phi_{\rm GW}|\lesssim10^{-3}$~rad in either case.}

We estimate the disk-induced apsidal phase-shift from $\langle a\rangle=53~r_s$ to $32~r_s$ to be $\delta\phi_{\dot\varpi_{\rm gas},\langle  a\rangle}{\lesssim}2\times10^{-5}$~rad, much smaller than the orbital phase-shift over the same range of semi-major axes. We measure this phase shift until $32~r_s$ because below this value the eccentricity decreases down to $e\lesssim0.01$ that is too small to reliably measure the gas-induced drift of the argument of periapsis.

We then compare the phase shifts at fixed time over the first $500$ orbits of the gas+$2$PN run (see \S~\ref{Sec:Postprocess}), both with and without GW emission. With GW emission, we find $\delta\phi^{\rm (GW)}_{\rm orb,t}\approx{-6\times10^{-6}}$~rad, and without GW emission $\delta\phi^{\rm (NoGW)}_{\rm orb,t}\approx{-7\times10^{-6}}$~rad. {Both agree within a factor of $\approx2$ with the estimate including the accreted mass, obtained from $\Delta\Omega/\Omega=\Delta M/2M-3\,\Delta a/2a$ with $\Delta M/M=\fedd t/(50~{\rm Myr})$ and $\Delta a/a=(8\xi_{\rm acc+grav}-3)\fedd t/(50~{\rm Myr})$, i.e.~$\delta\phi_t\approx-(6\xi_{\rm acc+grav}-5/2)\,\fedd\,\Omega_{\rm B}t^2/(50~{\rm Myr})\approx-4\times10^{-6}$~rad after $500$ orbits.}

\begin{table}[H]
\centering
\renewcommand{\arraystretch}{1.5}
{
    \begin{tabular}{|C|C|}
        \hline
        {\rm \bf Phase-shift}&{\bf Value~[rad]}\\
        \hline
        \hline
        \delta \phi^{\rm (GW)}_{\rm orb,\langle a\rangle}[53~r_s\rightarrow a_{\rm dec}] & {-2.4\times10^{-4}}\\
        \hline
        {\delta \phi^{\rm (GW)}_{\rm orb,\langle a\rangle}[53~r_s\rightarrow45~r_s]} & {+2.5\times10^{-4}}\\
        \hline
        \delta\phi_{\dot\varpi_{\rm gas,\langle a\rangle}}[53~r_s\rightarrow32~r_s] & {\lesssim}2\times10^{-5}\\
        \hline
         \delta \phi_{\rm GW,\langle a\rangle}[53~r_s\rightarrow a_{\rm dec}] & {-4.8\times10^{-4}}\\
        \hline
        \delta \phi^{\rm (GW)}_{\rm orb,t}[0\rightarrow500~P_{\rm B}] & {-6\times10^{-6}}\\
        \hline
        \delta \phi^{\rm (NoGW)}_{\rm orb,t}[0\rightarrow500~P_{\rm B}] & {-7\times10^{-6}}\\
        \hline
    \end{tabular}
}
\caption{Phase shifts measured from our simulations: gas-induced orbital {$\delta \phi^{\rm (GW)}_{\rm orb,\langle a\rangle}$ for the gas+$2.5$PN vs $2.5$PN simulations between $\langle a\rangle=53~r_s$ and the decoupling radius $\langle a_{\rm dec}\rangle=13~r_s$, {and its peak value at $\langle a\rangle=45~r_s$,} disk-induced binary precession phase-shift $ \delta\phi_{\dot\varpi_{\rm gas,\langle a\rangle}}$ between $\langle a\rangle=53~r_s$ and $\langle a\rangle=32~r_s$, gas-induced GW phase-shift  $\delta \phi_{\rm GW,\langle a\rangle}$ by doubling $\delta \phi^{\rm (GW)}_{\rm orb,\langle a\rangle}$, and two phase-shifts over fixed time with ($\delta \phi^{\rm (GW)}_{\rm orb,t}$) or without GWs ($\delta \phi^{\rm (NoGW)}_{\rm orb,t}$)} over the first $500$ orbits of the gas+$2$PN run.}
\label{table:dphi}
\renewcommand{\arraystretch}{1}
\end{table}

As often done in the literature, one can semi-analytically compute orbital phase shift \citep{Garg2022,Dittmann2023,Duffell2024} by linearly summing the rates of change of semi-major axis due to GWs {($\dot a_{\rm GW}$; \citealt{Peters1964})} and gas {($\dot a_{\rm gas}$), $\dot a=\dot a_{\rm GW}+\dot a_{\rm gas}$. For an equal-mass circular MBHB that is accreting material from a CBD, we can use the Newtonian parametrization \citep{Duffell2024} 

\begin{equation}
  \dot a_{\rm gas}\approx \frac{(8\xi_{\rm acc+grav}-3)\fedd}{50~\rm Myr}a,
\end{equation}
for the evolution of the semi-major axis driven by the gaseous torques. We can then calculate the integral, an initial $a_i=53r_s$ to a final $a_f=a_{\rm dec}$ semi-major axis, of the difference between the phase evolution with and without the gas contribution \citep[see, e.g.,][]{Derdzinski2021,Garg2022}. {We also include the accreted mass $\Delta M=\fedd M\,(t-t_i)/50~{\rm Myr}$ \citep[see also][]{Caputo2020,Garg2024b}, which increases both the GW-driven inspiral rate ($\dot a_{\rm GW}\propto M^3$) and the orbital frequency ($\Omega\propto M^{1/2}$) at a given separation, so that the analytical total dephasing can be written as}
\begin{equation}\label{Eq:integral}
    \delta\phi^{\rm analytical}_{{\rm orb},\langle a\rangle}=\int_{a_f}^{a_i}\frac{\Omega}{|\dot a_{\rm GW}|}\left[\frac{\dot a_{\rm gas}}{|\dot a_{\rm GW}|}-\frac{5}{2}\frac{\Delta M}{M}\right]da.
\end{equation}}
For our $10^6~\MSun$ binary, {using $\xi_{\rm acc+grav}\approx0.45$ and $\fedd\approx0.55$,} we obtain

\begin{align}\label{Eq:analytical}
    \delta \phi^{\rm analytical}_{\rm orb,\langle a\rangle}\approx\;&4\times10^{-15}~\fedd\,\Big[\big(8\xi_{\rm acc+grav}-3\big)\big(x_i^{6.5}-x_f^{6.5}\big)\\
    &{+\Big\{\frac{5}{8}\big(x_i^{6.5}-x_f^{6.5}\big)-\frac{13}{8}\,x_i^{4}\big(x_i^{2.5}-x_f^{2.5}\big)\Big\}}\Big]~{\rm rad}\nonumber\\
    \approx\;&{-1.2\times10^{-4}}~{\rm rad}\nonumber,
\end{align}}
{where $x\equiv a/r_s$. The first term is the usual estimate \citep[see, e.g.,][]{Duffell2024}, while the second and third terms come from the effect of the accreted mass on the GW inspiral rate, present only in the gas+GW simulation.
For $a_f\ll a_i$, the bracket becomes $(8\xi_{\rm acc+grav}-4)\,x_i^{6.5}$, which is negative for $\xi_{\rm acc+grav}<0.5$, as here. {For a constant $\fedd$,} the phase shift changes sign at $a\approx\lambda(\xi_{\rm acc+grav})\,a_i$, with $\lambda\approx0.59$ for $\xi_{\rm acc+grav}=0.45$, i.e.~at $\approx31.5~r_s$, close to the $\approx32~r_s$ we measure. {In our simulation $\fedd$ halves by then (Fig.~\ref{fig:fEdd}), which moves the change of sign in Eq.~\ref{Eq:integral} to $\approx33~r_s$. The radii of the change of sign and of the peak therefore depend on the accretion history and on $a_i$, so they are specific to our simulation rather than general features of accreting LISA binaries.} In Eq.~\ref{Eq:analytical}, the term $8\xi_{\rm acc+grav}$ is due to the gas torque, and the other terms to accretion. The phase shift peaks near $45~r_s$, where accretion starts to change the phase faster than the torque, and changes sign where the two contributions are equal ($\approx1.2\times10^{-3}$~rad each).}

{Fig.~\ref{fig:dphi_time} compares Eq.~\ref{Eq:analytical} with the measured phase shift. Without the terms in braces, it gives $\approx+2.1\times10^{-4}$~rad and misses the decline after $\sim40~r_s$, whereas with them it reproduces the turnover and the change of sign within a factor $\approx2.5$. Since the torque and accretion terms nearly cancel, the result is very sensitive to $\xi_{\rm acc+grav}$: a change of $0.05$ shifts the phase at $a_{\rm dec}$ by $\approx1.4\times10^{-4}$~rad, comparable to the difference from the simulation. Varying {a constant} $\fedd$ within the range of our simulation instead changes only the amplitude (gray band in Fig.~\ref{fig:dphi_time}). The GW phase shift, $|\delta\phi_{\rm GW,\langle a\rangle}|\lesssim5\times10^{-4}$~rad, would need SNR$\gtrsim8/|\delta\phi_{\rm GW,\langle a\rangle}|\approx16000$ to be detected by LISA \citep[see, e.g.,][]{Garg2022}, i.e.~a source at $z\lesssim0.1$ rather than at our fiducial $z=1$ (SNR$\approx1300$; \citealt{Garg2024b}).}

\section{Discussion and Conclusion}

We simulated the evolution of an equal-mass $10^6~\MSun$ quasi-circular MBHB embedded in a locally isothermal circumbinary disk. The binary orbit is described using {a} $2.5$PN expansion and therefore evolves under the effect of both GW emission and gaseous torques. We take the initial separation to correspond to {$1.5$} years before merger{, and evolve all our simulations for this time span, such that we can follow the binary inspiral until coalescence.} Following up on the work in Paper I \citep{Garg2025}, which focused on a thin disk ($H/R=0.03$), we {investigate here the effect of a disk with a larger aspect ratio $H/R=0.1$, so that the two studies bracket the plausible range of aspect ratios for CBDs around LISA MBHBs (see \S~\ref{S:Intro}).}

We measured the values of the gravitational and accretion torques onto the binary, finding overall positive values. This indicates that the gas torque is slowing down the inspiral in this configuration. 
{We find torque eigenvalues consistent with the ones reported in \citet{Duffell2024}. The gravitational torque reported here $\bar\xi_{\rm grav}=0.16$ is similar to the value shown by the steady state gravitational torque in Fig.~14 (upper panel) of \citet{Duffell2024}, divided by the accretion rate (lower panel in the same Fig.~14). The smaller value $\approx0.01$ shown for GIZMO in the upper panel of their Fig.~15 is instead a single value around the $250^{\rm th}$ orbit, and is therefore more affected by the variability of the torque than our average over hundreds of orbits.
We also find $\bar\xi_{\rm acc}\sim0.29$, compatible with the difference between the values of the accreted angular momentum and the fraction that goes into the spin of the MBHs reported in Table~2 of \citet{Duffell2024}.}

We found the disk to be able to follow the binary inspiral {down to the decoupling radius of $13~r_s$, i.e., about two days before coalescence}, and the minidisks to maintain their structure {at least down to $25~r_s$, i.e., about a month before merger.}

{As shown in \S~\ref{S:Mdot}, the lines in the accretion-rate variability near the binary orbital frequency ($\Omega_{\rm S}$) and twice it ($\Omega_{\rm D}$) are shifted to lower frequencies by, and therefore trace, the precession of the binary and of the disc cavity edge, respectively{, while the lump modulation ($\Omega_{\rm L}$) is also shifted by the cavity precession}. This makes our study useful for the correct identification of MBHB candidates \citep{Xin2026} from periodicities in time-domain surveys, such as the Vera Rubin Observatory's Legacy Survey of Space and Time (LSST; \citealt{LSST2019}) and the Roman telescope \citep{Haiman2023}, both now and especially in archival searches once LISA has tightly constrained the MBHB parameters \citep{Xin2024}.

As shown in Fig.~\ref{fig:fEddMR}, GW emission only weakens the modulation at $\Omega_{\rm S}$ through the circularization of the binary, while leaving the dominant lump modulation and that at $\Omega_{\rm D}$ unaffected. Hence, the mock light-curve catalogs \citep[see, e.g.,][]{DOrazio2024,Siwek2026} developed from fixed-orbit Newtonian simulations might be sufficient for the upcoming time-domain surveys for the $H/R=0.1$ CBD case, as long as a MBHB is more than a few years away from merger, once they account for this circularization and for the relativistic apsidal precession of the binary. To predict the electromagnetic detectability of MBHBs closer to merger, one would instead need to simulate the concurrent GW-driven evolution self-consistently, as we did in this work.

In \S~\ref{Sec:dphi}, we measured the GW phase shift $\delta\phi_{\rm GW,\langle a\rangle}$ accumulated as the binary inspirals in the presence of the gaseous disk. Between the initial separation and the decoupling radius, we find {$|\delta\phi_{\rm GW,\langle a\rangle}|\lesssim5\times10^{-4}$~rad, within a factor $\approx2.5$ of the semi-analytical estimate once the effect of the accreted mass on the GW-driven inspiral rate is included (Eq.~\ref{Eq:analytical}). The phase shift changes sign near $32~r_s$, where the accumulated effect of the accreted mass on the GW inspiral rate overtakes that of the positive gas torques{, which weaken as the accretion rate drops. This radius depends on the accretion history and on the initial separation, so it is specific to our setup}.} In contrast, Paper I \citep{Garg2025} found a larger disagreement with the semi-analytical approximation for a thinner disk around a moderately eccentric binary. Hence, for a circular binary in a warm disk the semi-analytical estimate can perhaps be trusted to within a factor of a few.} {The gas torques and GW emission appear to combine linearly, but accretion adds an effective coupling, since the accreted mass increases both the orbital frequency and the GW-driven inspiral rate. This effect, included in GW-phase models \citep{Caputo2020,Garg2024b} but usually neglected in torque-based estimates, changes $(8\xi_{\rm acc+grav}-3)$ into $(8\xi_{\rm acc+grav}-4)$ at late times (Eq.~\ref{Eq:analytical}). It can thus reverse the sign of the phase shift for $\xi_{\rm acc+grav}$ of order unity, as in warm disks, but is negligible for strongly negative torques, such as $\xi_{\rm acc+grav}\approx-20$ in Paper~I.}

There are some {notable} differences between the results of this work and the ones reported in Paper I{, despite starting at nearly the same binary separation.} In the previous study, the decoupling process was faster{, since a thinner disk has a longer viscous time and is therefore less able} to follow the binary inspiral.
In Paper I, the accretion torque magnitude was only $\sim 1\%$ of the gravitational torque, while in this study it is {almost} twice the gravitational torque. This difference is consistent with higher aspect ratio disks allowing more material to leak into the cavity, providing more accretion torque \citep{Ragusa2016,Franchini2022}.
We found that, while the {disk} was significantly aiding the inspiral in Paper I, owing to the dominant {negative} contribution from the {outer} CBD {and negligible ones from the short-lived minidisks and the accretion torque, here the total gas torque is positive, because the positive torque from the minidisks outweighs the negative one from the outer CBD, and the large positive accretion torque adds to it.} This result is consistent with the literature on the effect of the disk thickness on the torque sign \citep{Tiede2020,Franchini2022}.

\subsection{Caveats}

Similarly to Paper I, the main caveat of this work is the assumption that a steady-state non-GW CBD configuration can be used as an initial condition for the GW+gas simulation at the separations explored. {Simulations following the} evolution of MBHBs from $\sim0.1$ pc down to $\sim10^{-5}$ pc, and {which would inform the} expected CBD configuration, are {currently} prohibitively expensive in terms of computational cost. Therefore the approach adopted here remains to date the most feasible way to include GW emission in hydrodynamics simulations to infer predictions on LISA observables.

Another caveat is that we assumed the transport in the disk to be driven by viscosity, parametrized with the Shakura-Sunyaev approximation. 
CBDs that host near-merger LISA MBHBs might {consist of} hot, magnetized plasma with magneto-hydrodynamical turbulence supported by the magneto-rotational instability \citep[see, e.g.,][]{Murphy2015,Jiang2025,Tiwari2025}.
The effect of such disks on the binary evolution {remains poorly} understood. A recent semi-analytical work \citep{Garg2026} showed that the resulting gas-induced phase shifts can differ significantly from the laminar flow case. 
However, again due to the high computational cost, our current approach is the most affordable option {to self-consistently evolve the binary under both GWs and gas torques during the last few years before merger.} 

{For the same reason, we do} not {consider} higher-order general relativistic corrections or full general relativity, typically used in electromagnetic studies \citep[see, e.g.,][]{Avara2024,Ressler2025}. Integrating a handful of orbits a few days before merger in full general relativity is the only viable way to avoid prohibitive computation times but does not provide useful information on the dephasing that can be observed in the LISA band, which requires simulating the binary at much larger separations and for several {thousands of} binary orbits. 
{These approximations become less accurate near the end of the inspiral: at the decoupling radius the binary orbital velocity reaches $v/c\approx0.2$, so higher-order PN terms and relativistic effects on the gas, which we treat with Newtonian gravity, may no longer be negligible, especially in the minidisks. However, the gas-induced phase shift is largest near $45~r_s$, where $v/c\approx0.1$.}
{We initialize our binary about $1.5$~yr before merger and measure the phase shift between simulations with and without gas, but with the same PN order. Therefore, higher-order terms should cancel out, to leading order, in the gas-induced phase difference, so that our measurement remains reliable.} {Finally, we compare to a vacuum binary with the same initial mass. In a LISA analysis, the binary mass is a free parameter, so part of the accreted-mass contribution would be absorbed into the inferred mass. Quantifying this, and other degeneracies with the binary parameters, requires a parameter-estimation study beyond the scope of this work.}

In conclusion, our work {supports the importance of including concurrent GW emission in CBD hydrodynamical simulations of MBHBs in their last years before merger and provides a benchmark for future studies.} It will also help improve the modeling of the imprint of gas on LISA sources and {of} the expected emission signatures for electromagnetic surveys, leading to better multimessenger synergies and strategies.

\section*{Acknowledgments}
We acknowledge useful discussions with {Daniel D'Orazio and} Andrew MacFadyen. {We thank the anonymous referee for their helpful comments that improved this work. We further acknowledge the use of the NYU IT HPC Torch cluster.} MG acknowledges support from the Swiss National Science Foundation (SNSF) Postdoc Mobility fellowship P500-2\_235363. The authors also acknowledge {use of the Claude Sonnet 5 \citep{Anthropic2026} for grammar correction and debugging analysis scripts,} and NumPy \citep{harris2020array}, Matplotlib \citep{Hunter2007}, Pandas \citep{Pandas}, and SciPy \citep{SciPy2020} Python packages.

\bibliography{TorqueLISAII}
\bibliographystyle{aasjournal}
\normalsize

\appendix
\setcounter{figure}{0}                       
\renewcommand\thefigure{A\arabic{figure}}  

\section{Resolution study}\label{App:Res}

In this section, we perform a resolution study for the gas+$2.5$PN setup by considering $100$ orbits each for the low-resolution (LR) with $\Delta x[3a]={3.6~r_s}$, mid-resolution (MR) with $\Delta x[3a]={2.1~r_s}$, and high-resolution (HR) with $\Delta x[3a]={1.2~r_s}$ simulation. After $100P_{\rm B}$, the binary shrinks {from $\langle a\rangle\approx$}$53~r_s$ {to} $51.9~r_s$. 

\begin{figure*}
    \centering
    \includegraphics[width=0.9\linewidth]{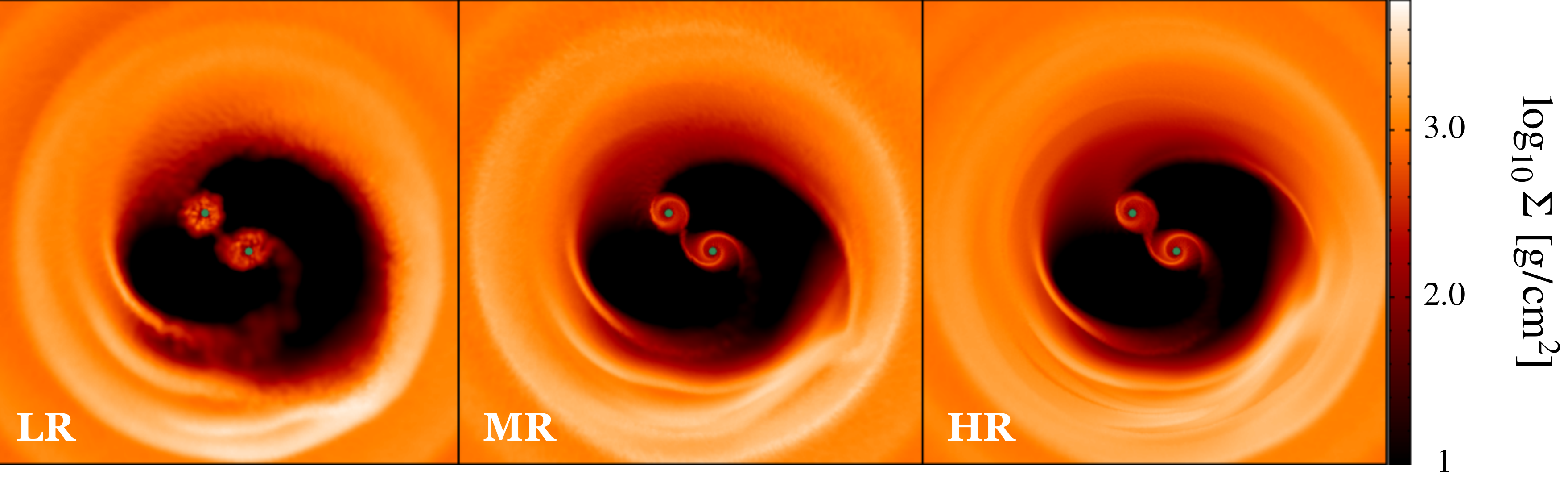}
    \caption{Same as Fig.~\ref{fig:disk} but comparing column densities at $a=53~r_s$ for the gas+$2.5$PN setup  at different resolutions: LR (left panel), MR (middle panel), and HR (right panel).}
    \label{fig:disk_comp}
\end{figure*}

In Fig.~\ref{fig:disk_comp}, we show the disk morphology at the start of the simulation for each resolution. We can see that the minidisks and gas streams become much more defined in the MR run compared to the LR one.
In the HR run, the density waves at the inner edge of the cavity are more clearly defined.

\begin{figure}[h]
    \centering
    \includegraphics[width=0.465\textwidth]{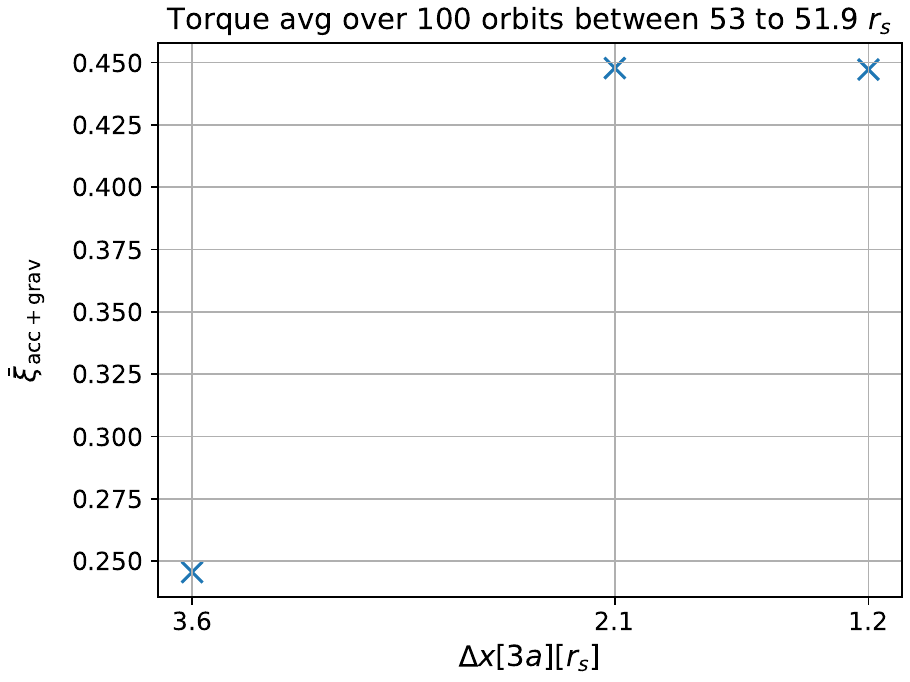}
    \includegraphics[width=0.45\textwidth]{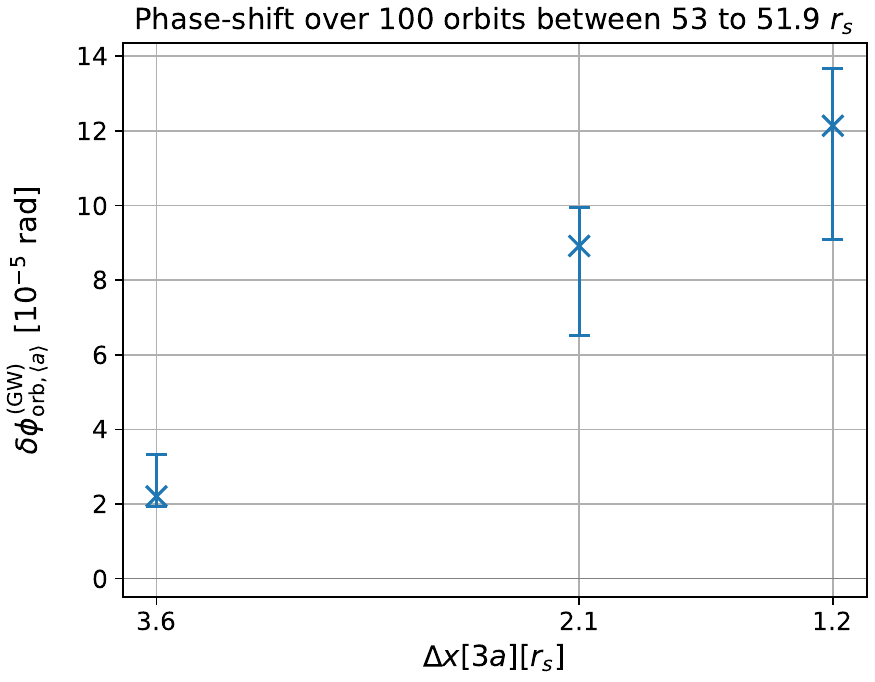}
    \caption{The mean dimensionless torque parameter $\bar\xi_{\rm grav+acc}$ {(left panel) and the gas-induced orbital phase shift at fixed $\langle a\rangle$ between $53$ and $51.9~r_s$ (right panel; error bars: conservative uncertainty from varying the window end between $52.2$ and $51.7~r_s$),} {over $100$ orbits starting near $\langle a\rangle=53~r_s$,} computed from the gas+$2.5$PN simulations for three different resolutions: LR with $\Delta x[3a]={3.6~r_s}$, MR with $\Delta x[3a]={2.1~r_s}$, and HR with $\Delta x[3a]={1.2~r_s}$.}
    \label{fig:Torque_res}
\end{figure}

In Fig.~\ref{fig:Torque_res}, we compare the total torque value $\bar\xi_{\rm acc+grav}$ across the three resolutions. 
The torque measured in the LR run is $\sim25\%$ lower than that in the MR, making the LR run unsuitable for our purposes. The difference in torque between the MR and HR run is instead $\lesssim1\%$. 

We also {compute} $\delta \phi^{\rm (GW)}_{\rm orb,{\langle a\rangle}}$ {for each resolution (Fig.~\ref{fig:Torque_res}, right panel). It increases with resolution, but the change halves with each refinement ($6.7$ and $3.2\times10^{-5}$~rad), so MR underestimates the converged value by only $\lesssim40\%$. The difference between MR and HR is tiny, only $\approx5\times10^{-8}$ of the $\approx670$~rad of orbital phase accumulated over these orbits}. Hence, at MR the disk is sufficiently resolved for our purposes.

{To test convergence in time, we also repeated the gas+$2.5$PN MR simulation with the MBH timestep halved to $\approx1.1$~s, and compared it with a $2.5$PN run at the same timestep. Over $80$ orbits, halving the timestep changes the phase shift only from $7.48\times10^{-5}$~rad to $7.47\times10^{-5}$~rad, i.e.~by less than $1\%$, so the phase shift is converged in time.}

\setcounter{figure}{0}                       
\renewcommand\thefigure{B\arabic{figure}}  

\section{Additional accretion rate information}\label{App:Spec}

\begin{figure}
    \centering
    \includegraphics[width=0.5\textwidth]{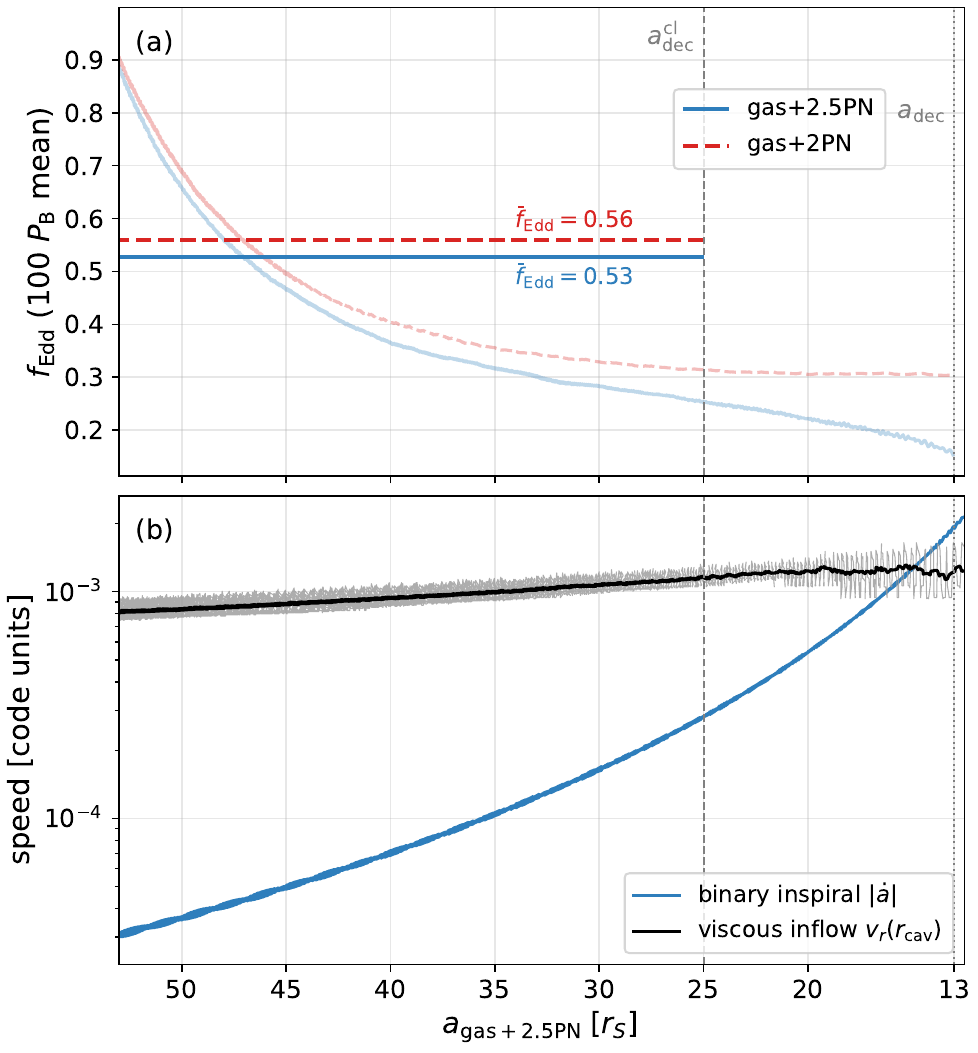}
    \caption{{(a)} Same as Fig.~\ref{fig:Gastorque} but for $\fedd$, extended to $a_{\rm dec}\approx13~r_s$ {(dotted line), with means over $a\geq a^{\rm cl}_{\rm dec}$ (thick lines)}. Near $a_{\rm dec}$ the gas+$2.5$PN averaging window is shortened so that $a$ changes by $\leq10\%$ across it. The gas+$2$PN accretion rate declines as our finite disk drains, while the gas+$2.5$PN one falls below it, to $\approx80\%$ at $a^{\rm cl}_{\rm dec}$ (dashed line) and $\approx50\%$ at $a_{\rm dec}$, as the binary decouples. {(b) Binary inspiral speed $|\dot a|$ and viscous inflow speed at the cavity edge $v_r(r_{\rm cav})=3\nu/2r_{\rm cav}$ (gray: raw; black: smoothed), in code units. The binary decouples once $|\dot a|$ exceeds the raw $v_r$ for good, at $a_{\rm dec}$ (Eq.~\ref{eq:adec}).}}
    \label{fig:fEdd}
\end{figure}

{In Fig.~\ref{fig:fEdd}, we show the evolution of the Eddington ratio for our two CBD runs{, together with the binary inspiral speed and the viscous inflow speed at the cavity edge, whose crossing defines $a_{\rm dec}$}.}

\begin{figure*}
    \centering
    \includegraphics[width=0.9\linewidth]{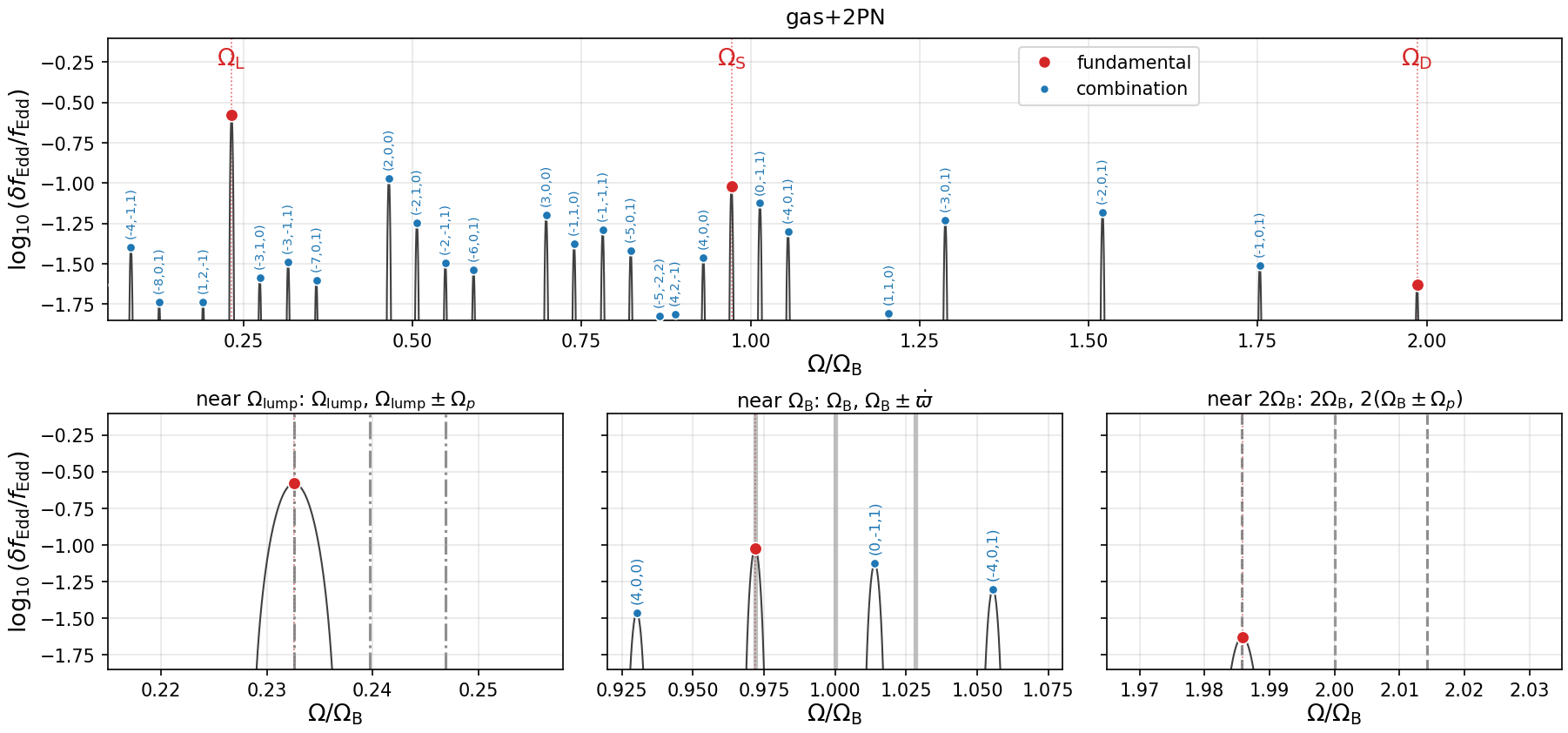}
    \caption{Similar to Fig.~\ref{fig:fEddMR}, but the periodogram of $\fedd$ for the gas+$2$PN simulation computed over $500$ orbits (frequency resolution $0.002~\Omega_{\rm B}$). Top panel: red points mark the three fundamental frequencies $\{\Omega_{\rm L},\Omega_{\rm S},\Omega_{\rm D}\}$ as defined in Fig.~\ref{fig:fEddMR} and the rest, denoted by blue points, are combinations of these three frequencies, labeled by the tuple $(n_1,n_2,n_3)$ corresponding to the frequency identified as $n_1\Omega_{\rm L}+n_2\Omega_{\rm S}+n_3\Omega_{\rm D}$. {Bottom-left panel: zoom near the lump frequency, with dash-dotted gray lines at $\Omega_{\rm L}=\Omega_{\rm lump}-\Omega_p$, $\Omega_{\rm lump}$ (the $m=1$ pattern speed of the lump measured from the snapshots), and $\Omega_{\rm lump}+\Omega_p$.} {Bottom-middle} panel: zoom near $\Omega_{\rm B}$ with solid gray lines at $\Omega_{\rm S}=\Omega_{\rm B}-\dot\varpi$, $\Omega_{\rm B}$, and $\Omega_{\rm B}+\dot\varpi$, the latter two carrying no power above the floor. Bottom-right panel: zoom near $2\Omega_{\rm B}$ with dashed gray lines at $\Omega_{\rm D}=2(\Omega_{\rm B}-\Omega_p)$, $2\Omega_{\rm B}$, and $2(\Omega_{\rm B}+\Omega_p)$, again the latter two carrying no detectable power.}
    \label{fig:Periodogram}
\end{figure*}

{In Fig.~\ref{fig:Periodogram}, we show the periodogram of $\fedd$ for the gas+$2$PN run over the same $500$ orbits as in Fig.~\ref{fig:fEddMR}, which gives the fractional amplitude of the $\fedd$ modulation at each frequency.}

\begin{figure*}
    \centering
    \includegraphics[width=0.9\linewidth]{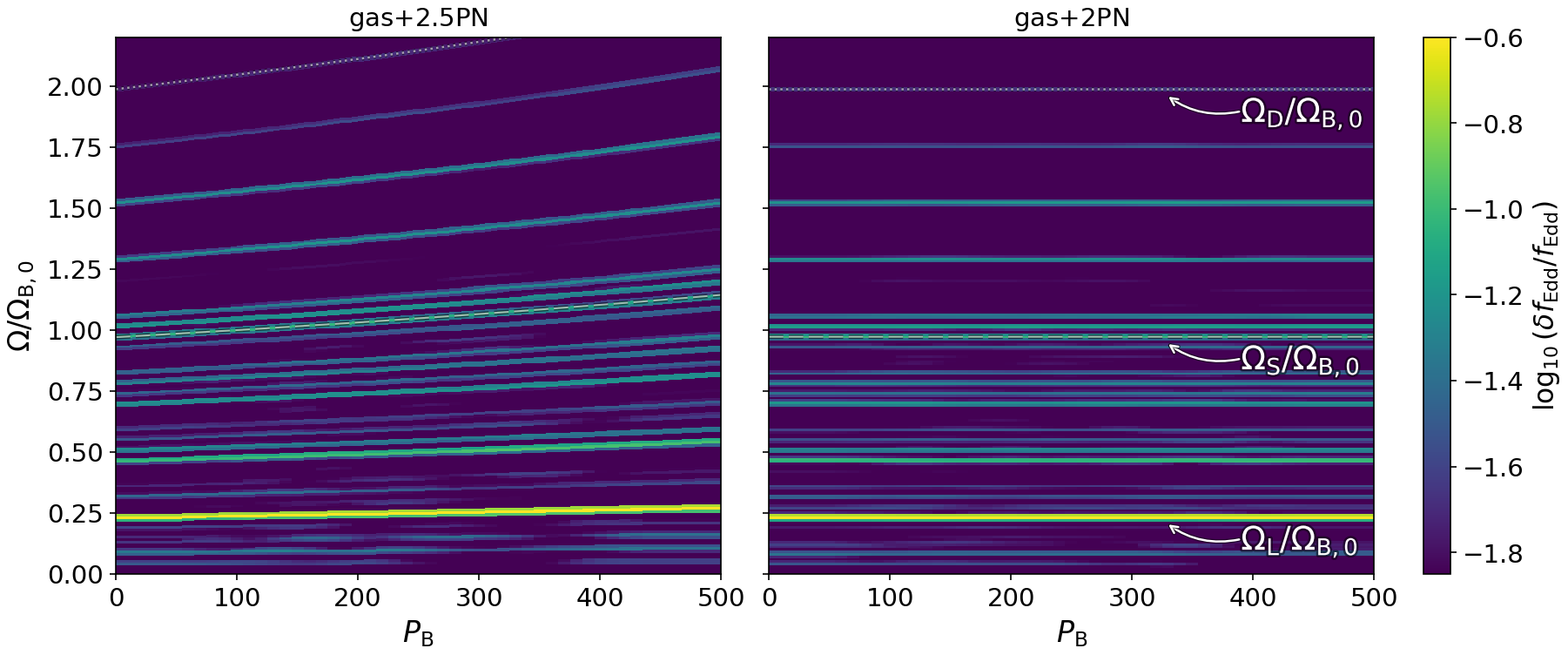}
    \caption{Same as Fig.~\ref{fig:fEddMR} but now $\Omega$ normalized by the initial orbital angular frequency $\Omega_{\rm B,0}$.}
    \label{fig:Spec_initial}
\end{figure*}

In Fig.~\ref{fig:Spec_initial}, we show the spectrograms similar to Fig.~\ref{fig:fEddMR} but now for frequencies $\Omega$ normalized by the initial frequency $\Omega_{\rm B,0}$. Due to GWs, there is a drift in the location of peak frequencies during the orbital evolution, as instantaneous $\Omega_{\rm B}$ is increasing due to the binary shrinkage. A similar behavior is reported by \citet{Franchini2024}.

\setcounter{figure}{0}
\renewcommand\thefigure{C\arabic{figure}}

\end{document}